# A density functional theory with correct long-range asymptotic behavior

Roi Baer

*Depart. of Physical Chemistry and the Lise Meitner Minerva-Center for Computational Quantum Chemistry, the Hebrew University of Jerusalem, Jerusalem 91904 Israel.*

Daniel Neuhauser

*Dept. of Chemistry and Biochemistry, University of California at Los Angeles, CA 90095, USA*

We derive an exact representation of the exchange-correlation energy within density functional theory (DFT) which spawns a class of approximations leading to correct long-range asymptotic behavior. In what amounts to be the simplest approximation, we obtain an electronic structure theory that combines a new local correlation energy (based on Monte-Carlo calculations applied to the homogeneous electron gas (HEG)) and a combination of local and explicit *long-ranged exchange*. The theory is applied to several 1$^{st}$ row atoms and diatomic molecules where encouraging results are obtained: a good description of the chemical bond at the same time allowing for bound anions, reasonably accurate affinity energies and correct polarizability of an elongated hydrogen chain. Further stringent tests of DFT are passed: the ratio of ionization potential and highest orbital energy is close to unity and the atomic charge of two distant hydrogen atoms under large bias is quantized (integer).

Density functional theory (DFT)[1, 2] is an in-principle exact approach to molecular electronic structure furnishing a starting point for useful approximations. The local density (LDA)[2] and local spin-density (LSDA) approximations form a surprisingly well-balanced, robust and reasonably accurate theories for ground-state properties of molecules, solids and even metals[3-5]. The generalized gradients approximation (GGA)[6, 7] improves accuracy further and in finite systems hybrid functional approximations (HFA)[8, 9] are often even more powerful. Despite their great success, present functionals fail to account for anions and processes involving long-range charge transfer. For example, the polarizability of elongated molecules is strongly exaggerated[10]. These problems are associated with the existence of self-interaction in the local functionals[11-15], and exchange-correlation (XC) potentials derived from them exhibit spurious exponential decay at large distances from the molecule or solid surface.[13, 16]

A particularly useful method for self-interaction correction (SIC) was suggested by Perdew and Zunger[13] (PZ), and has been successfully applied to anions and computation of electron affinities[17]. Recently, a systematic study of this method was performed[18]. The theoretical difficulties with this method were overcome by the optimized effective potential (OEP)[19], recently further developed in a form that can be applied to systems beyond atoms[20, 21]. SIC-OEP predicts correct polarizability and electron affinity where LSDA/GGA fail[21].

One well-known "type" of density functional theory not having the problem of self-interaction is the Hartree-Fock theory (HFT). In HFT, the interplay between exchange and Hartree terms results in exact self-interaction removal. HFT however, is not sufficiently accurate because it neglects correlation energy. Simply adding a local density correlation energy functional, the so-called Kohn-Sham HFT (KSHFT), to the explicit orbital exchange in HFT is unsatisfactory when compared with LSDA because the exact exchange disrupts a delicate cancellation of errors existing in LSDA[3].

This paper describes a new approach to self-interaction correction and long range behavior of DFT approximations. We derive a new exact representation of the correlation energy that *endorses* approximations with correct asymptotic behavior. A simple approximation using this theory is then shown to yield a new useful functional, exhibiting good description of chemical bonds while allowing for stable anions and correct polarizability of elongated molecules. The new functional is better suited for use in confined systems, where the HEG correlation energy is usually too large by about 100%. An account of the theory, followed by a demonstration on a few representative systems is given below.

We consider a system of $N$ electrons, with the Hamiltonian $\hat{H} = \hat{T} + \hat{V} + \hat{U}$ where $\hat{T} = \sum_{i=1}^{N}\left(-\frac{1}{2}\nabla_i^2\right)$ is the kinetic energy (we use atomic units, $e^2/4\pi\varepsilon_0 = \hbar = \mu_e = 1$) $\hat{V} = \sum_{i=1}^{N} v(\hat{\mathbf{r}}_i)$ the external potential and $\hat{U} = \frac{1}{2}\sum_{i\neq j=1}^{N_e} u(\hat{r}_{ij})$ (where $\hat{r}_{ij} = |\mathbf{r}_i - \mathbf{r}_j|$) is the interaction, with $u(r)$:

$$u(r) = \frac{1}{r}. \qquad (1)$$

In DFT, all expectation values are functionals of the ground-state density $n(\mathbf{r})$. The energy functional[2]

$$E_v[n] = T_s[n] + \int v(\mathbf{r})n(\mathbf{r})d^3r + E_H[n] + E_{XC}[n] \qquad (2)$$

is the basic functional of Kohn-Sham DFT since by minimizing it one maps the interacting electrons onto non-interacting fermions obtaining the ground-state density and energy. In Eq. (2), $T_s[n]$ is the non-interacting kinetic energy,

$$E_H = \frac{1}{2}\int n(\mathbf{r})n(\mathbf{r}')u(|\mathbf{r}-\mathbf{r}'|)d^3r d^3r' \qquad (3)$$

is the Hartree energy and

$$E_{XC}[n] = T[n] - T_s[n] + U[n] - E_H[n] \quad (4)$$

the exchange-correlation energy (XCE), where $T[n]$ is the kinetic energy and $U[n]$ the potential energy.

The XCE is of course practically impossible to determine exactly. It can also be written, using a straightforward extension of the adiabatic connection theorem[22-24]. For this, we consider a family of systems continuously parameterized by $0 \leq \gamma < \infty$. All systems have the same ground-state density $n$ but the particles in each system interact via a different descreened two-body interaction:

$$u_\gamma(r) = \frac{1 - e^{-\gamma r}}{r}, \quad (5)$$

For large inter-particle distances $\gamma r \gg 1$, the particles of system $\gamma$ repel just like electrons, but at short distances the repulsion is moderated and non-singular. Each system $\gamma$ has a unique ground-state wavefunction $\Psi_\gamma$ (assuming v-representability). At $\gamma = \infty$ $\Psi_\infty$ is the ground-state wavefunction of the Coulomb interacting system and at $\gamma = 0$, corresponding to non-interacting particles, $\Psi_0$ is a Slater determinant of $N$ spin-orbitals. The adiabatic-connection theorem now states:

$$E_{XC}[n] = \int_0^\infty \langle \Psi_{\gamma'} | \hat{W}_{\gamma'} | \Psi_{\gamma'} \rangle d\gamma' - E_H[n], \quad (6)$$

where $\hat{W}_\gamma = \frac{1}{2} \sum_{i \neq j} w_\gamma(\hat{r}_{ij})$, $w_\gamma(r) = e^{-\gamma r}$.

Evaluating the XCE Eq. (6) is once again next to impossible. Yet, a simple approximation already leads to a meaningful theory: assume that in Eq. (6) $\Psi_\gamma$ is replaced by $\Psi_0$. Under this approximation, the integral can be performed and yields the HFT exchange-energy. The next-simple approximation is to assume $\Psi_{\gamma'} = \Psi_0$ for $\gamma' < \gamma$ and $\Psi_{\gamma'} = \Psi_\infty$ for $\gamma' > \gamma$, for some $0 < \gamma < \infty$, giving:

$$\int_0^\infty \langle \Psi_{\gamma'} | \hat{W}_{\gamma'} | \Psi_{\gamma'} \rangle d\gamma' \approx \langle \Psi_0 | \hat{U}_\gamma | \Psi_0 \rangle + \langle \Psi_\infty | \hat{Y}_\gamma | \Psi_\infty \rangle \quad (7)$$

where: $\hat{U}_\gamma = \frac{1}{2} \sum_{i \neq j} u_\gamma(\hat{r}_{ij})$ and $\hat{Y}_\gamma = \frac{1}{2} \sum_{i \neq j} y_\gamma(\hat{r}_{ij})$, with $y_\gamma(r) = e^{-\gamma r}/r$ the Yukawa potential. The first term on the right of Eq. (7) is simply the HFT potential energy $\langle \Psi_0 | \hat{U}_\gamma | \Psi_0 \rangle \equiv E_H^\gamma[n] + K_X^\gamma[n]$. Here $E_H^\gamma[n]$ is defined by Eq. (3) with the Coulomb potential $u(r)$ replaced by the descreened potential $u_\gamma(r)$ and

$$K_X^\gamma[n] = \frac{1}{2} \int |P[n](\mathbf{r}, \mathbf{r}')|^2 u_\gamma(|\mathbf{r} - \mathbf{r}'|) d^3r d^3r' \quad (8)$$

is the corresponding exchange energy. In Eq. (8) $P[n](\mathbf{r}, \mathbf{r}')$ is the density matrix of non-interacting electrons having density $n$. From Eq. (4), (6) and (7), the error, i.e. the difference between the exact and approximate XCE is:

$$\Delta E_{XC}^\gamma = \left[ T[n] + \langle \Psi_\infty | \hat{U}_\gamma | \Psi_\infty \rangle \right] - \left[ T_s[n] + \langle \Psi_0 | \hat{U}_\gamma | \Psi_0 \rangle \right] \quad (9)$$

We now study two limits of the approximation. For $\gamma = 0$, $U_\gamma = 0$ so the error is $\Delta E_{XC}^{\gamma=0} = T[n] - T_s[n]$, a manifestly *non-negative* quantity[4], while for $\gamma \to \infty$ $U_\gamma \to U$ and so $\Delta E_{XC}^\gamma \to E_C$, where $E_C$ is the correlation energy, a manifestly *non-positive* quantity. Thus, assuming continuity, we conclude that there must always exist $0 \leq \gamma \leq \infty$ (in general depending on the density) for which the approximation (7) is *exact*. This allows us to assert that the energy functional $E_v$ can be written accurately as:

$$E_v[n] = T_s[n] + E_{ext}[n] + E_H[n] + K_X^\gamma[n] + E_{XC}^\gamma[n] \quad (10)$$

where:

$$E_{XC}^\gamma[n] = \langle \Psi_\infty | \hat{Y}_\gamma | \Psi_\infty \rangle - \frac{1}{2} \int n(\mathbf{r}) n(\mathbf{r}') y_\gamma(|\mathbf{r} - \mathbf{r}'|) d^3r d^3r' \quad (11)$$

This exact representation of the correlation energy (i.e. the sum of the last two terms in Eq. (10)) is a starting point for approximations, concentrating on simplified forms for the functionals $\gamma[n]$ and $E_{xc}^\gamma[n]$. The important feature is that this approach allows to naturally introduce the explicit exchange functional $K_X^\gamma[n]$. This functional produces a potential which at large distances ($r\gamma \gg 1$) corrects the spurious self interaction produced by the Hartree-potential derived from $E_H[n]$. Thus, $K_X^\gamma[n]$ heals the ailments of local and generalized gradient density approximations associated with the spurious long-range self interaction.

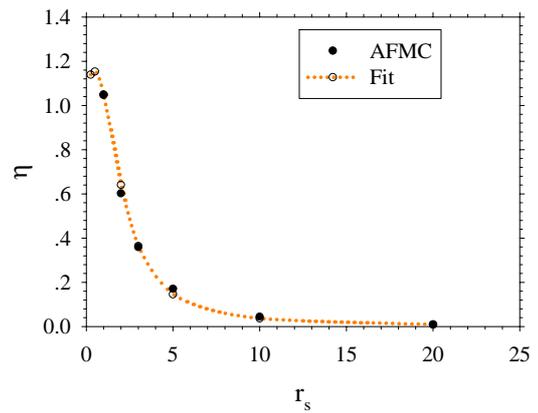

Figure 1: The ratio $\eta = \varepsilon_C^\gamma / \varepsilon_C^0$ as a function of $r_s$ for $\gamma = 1 a_0^{-1}$. Shown, the AFMC results (filled dots) and the analytical fit (dotted curve and white circles).

Let us now make the simplest and perhaps crudest approximation: assume $\gamma$ is *completely independent* of $n$. We now show that this leads to a useful molecular electronic-structure theory. For definiteness, we further assume $\gamma = 1 a_0^{-1}$. This value is arbitrary, and we have *not* attempted to optimize it at this stage (an optimization against a benchmark database will most likely improve the results shown here). Additionally, the energy of Eq. (11) is approximated as a local-density functional:

$$E_{XC}^\gamma [n] \approx \int \varepsilon_{XC}^\gamma (n(\mathbf{r})) n(\mathbf{r}) d^3 r, \qquad (12)$$

where $\varepsilon_{XC}^\gamma (n)$ is the XCE per particle of Eq. (11) for a HEG. For convenience, we write this function as follows:

$$\varepsilon_{XC}^\gamma (n) = \varepsilon_X^\gamma (n) + \varepsilon_C^\gamma (n) \qquad (13)$$

Where $\varepsilon_X^\gamma (n)$ is the analytical local screened exchange in a HEG given in [25].

The function $\varepsilon_C^\gamma (n)$ was evaluated numerically for the HEG using the shifted-contour auxiliary field Monte-Carlo (SCAFMC) method[26, 27], performed with plane-waves. In this preliminary account we made no attempt to fully converge to the infinite cell size limit and the statistical error is estimated to be ~10%. A more rigorous calculation and a full account of the details of the calculation will be published elsewhere. We calculated $\varepsilon_C^\gamma (n)$ (with $\gamma = 1 a_0^{-1}$) at several densities given by $r_s = 1, 2, 3, 5, 10, 20$. For convenience of application, the results are expressed in terms of the ratio $\eta = \varepsilon_C^\gamma / \varepsilon_C^0$ where $\varepsilon_C^0$ is the (usual) full correlation energy for the HEG (parameterized in any DFT code).

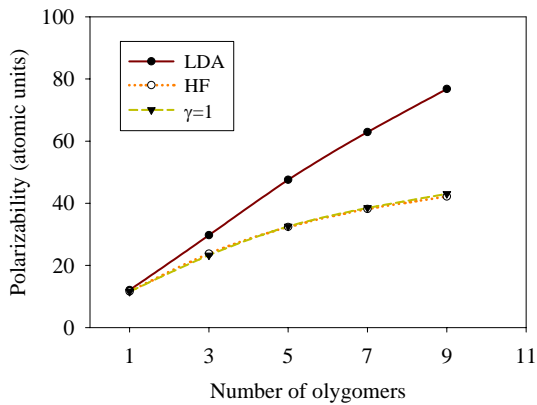

Figure 2: Polarizability of a linear Hydrogen chain, as a function of its length, calculated with LDA, HFT and present functional.

In the low density limit ($r_s \to \infty$) $\eta$ approaches $A/[\gamma r_s a_0]^2$ (with $A = 3.4602$) so we fitted the SCAFMC results to the following functional form:

$$\eta(r_s) \equiv \frac{\varepsilon_C^{\gamma=1}(n)}{\varepsilon_C^0(n)} = \frac{A}{C_0 + C_1 r_s + r_s^2}, \qquad (14)$$

finding $C_0 = 3.2$, $C_1 = -0.9$. The form of the ratio as a function of $r_s$ is given in Figure 1. For almost all densities, except the very high regime, $\eta$ is smaller than 1, decaying to zero as $r_s \to \infty$. Thus, $\varepsilon_{Xc}^\gamma$ is much smaller than $\varepsilon_{XC}^0$ as we know it should be for confined systems. It is interesting that the SCAFMC results show that $\eta$ is somewhat greater than 1 in the high-density limit. This result has to be further checked by increasing the accuracy of the SCAFMC calculation, a feat left for future investigations. For the purpose of demonstration, we used the ratio $\eta$ also for the spin-polarized correlation energy.

The functional of Eq. (10) contains a long-range portion of explicit exchange $K_X^\gamma [n]$, an a complimentary local density exchange and correlation functionals based on $\varepsilon_X^\gamma (n)$ and $\varepsilon_C^\gamma (n)$. In principle, a KS application of the functional requires an OEP approach[19]. To circumvent this complication, we minimized the energy $E_v$ as a functional of orbitals, instead of the density. Since both approaches are variational they are known to yield very similar orbitals and energies[19]. We now show this functional furnishes balanced molecular electronic structure: it has built-in correct long-range behavior and good description of the chemical bond. We performed several calculations done with a plane-waves code, using norm-conserving pseudopotentials[28] and a LSDA parameterization of the HEG correlation energy[29]. All results are fully self-consistent and converged with respect to cell size and grid spacing.

Table 1: Atomization energy, bond length and vibrational frequency and atomic affinity energy, comparing experimental data to the $\gamma = 1$ approximation and the LSDA.

| Atom (X) | $\Delta E (X_2)$ kcal/mol | $R_e (X_2)$ Angstrom | $\omega_e (X_2)$ cm$^{-1}$ | EA (X) eV |
|---|---|---|---|---|
| C | Exp. 145[30] $\gamma=1$ 132 LSDA 169 | 1.242[31] 1.27 1.25 | 1855[31] 1810 1770 | 1.26[32] 1.33 |
| N | Exp. 225[33] $\gamma=1$ 223 LSDA 256 | 1.098[31] 1.06 1.089 | 2358[31] 2540 2520 | Unstable Unstable |
| O | Exp. 118[33] $\gamma=1$ 122 LSDA 167 | 1.208[31] 1.14 1.20 | 1580[31] 1640 1590 | 1.46[32] 1.50 |
| F | exp. 37[33] $\gamma=1$ 35 LSDA 75 | 1.412[31] 1.330 1.390 | 916[31] 1200 1000 | 3.40[32] 3.73 |

We first discuss the computed polarizability of linear chains of hydrogen atoms[10], shown in Figure 2, using LDA, HFT and the new functional. The HFT results are known to be similar to accurate wave function methods[21], and the new functional gives essentially identical results to HFT, in contrast to LSDA which greatly overestimates polarizability.

Chemical bonds and atomic electron affinities are reasonably well described by the new functional as shown for several examples in Table 1, compared with experiments. Comparing to LSDA, it is seen that atomization energies are greatly improved, vibrational frequencies have comparable errors but bond lengths are worse. Atomic electron affinities are reproduced well with the maximal error of 0.34eV for F. An additional test is in the equality of $\varepsilon_H$ and $IP$, the former is the energy of the highest occupied level and $IP$ is the ionization potential[16]. For the atoms considered in Table 1, LSDA yields a poor result $\left(IP/\varepsilon_H\right)_{LDA} \approx 1.8$, while the new functional yields a good value, $\left(IP/\varepsilon_H\right)_{\gamma=1} \approx 1.05$. Another stringent test for DFT is the "derivative discontinuity" property: a weakly coupled molecule or atom must have an *integer* number of electrons [12]. We checked this by considering two distant hydrogen atoms under a bias $v_{bias}$. LSDA erroneously shows a continuous charge transfer between the atoms as $v_{bias}$ is increased. The new functional describes the correct physics: when $v_{bias} < IP - EA$ no charge is transferred while a complete electron is transferred once $v_{bias} > IP - EA$.

Summarizing, we developed an exact representation of the XC energy, constructing a new approximate functional exhibiting correct long-range behavior. The function $v_\gamma(r)$ (Eq. (5)) is not fundamental: any interaction that smoothly turns off the Coulomb potential "from the inside out" leads to a similar theory. Recently the form $erf(\gamma r)$ (replacing the numerator in Eq. (5)) was used[34] in an exchange-only calculation. Related approaches to ours[35, 36] separated the interaction to long-range and short-range parts. Our method is different, since the new exchange functional is matched by a new correlation functional.

Future work will address functional developments and new applications. Application within a Gaussian basis set is a first step, taking core electrons into account. Next we will improve the Monte Carlo calculations, seeking the infinite cell limit and parameterize a spin-polarized $\varepsilon_C^\gamma(n)$. Optimization of $\gamma$ is important as well, as searching for efficacious forms of $\gamma[n]$. Applications await, computing molecular response, excitations, especially for Rydberg states using time-dependent DFT. In molecular conductance[37] self-repulsion removal may be extremely important, as they are in chemistry at metallic surfaces[38] and strong lasers[39].

Support by Israel Science Foundation and NSF is acknowledged.

[1] P. Hohenberg and W. Kohn, Phys. Rev. **136**, B864 (1964).
[2] W. Kohn and L. J. Sham, Phys. Rev **140**, A1133 (1965).
[3] R. G. Parr and W. Yang, *Density Functional Theory of Atoms and Molecules* (Oxford University Press, Oxford, 1989).
[4] R. M. Dreizler and E. K. U. Gross, *Density Functional Theory: An Approach to the Quantum Many Body Problem* (Springer, Berlin, 1990).
[5] W. Kohn, A. D. Becke, and R. G. Parr, J. Phys. Chem. **100**, 12974 (1996).
[6] D. C. Langreth and M. J. Mehl, Physical Review B **28**, 1809 (1983).
[7] A. D. Becke, Phys Rev A **38**, 3098 (1988).
[8] A. D. Becke, J. Chem. Phys. **98**, 1372 (1993).
[9] C. Adamo and V. Barone, J. Chem. Phys. **110**, 6158 (1999).
[10] P. L. de Boeij, F. Kootstra, J. A. Berger, R. van Leeuwen, and J. G. Snijders, J. Chem. Phys. **115**, 1995 (2001).
[11] E. Fermi and E. Amaldi, Accad. Ital. Rome **6**, 119 (1934).
[12] J. P. Perdew, R. G. Parr, M. Levy, and J. L. Balduz, Phys. Rev. Lett. **49**, 1691 (1982).
[13] J. P. Perdew and A. Zunger, Phys. Rev. B **23**, 5048 (1981).
[14] A. Dreuw, J. Wiesman, and M. Head-Gordon, J. Chem. Phys. **119**, 2943 (2003).
[15] D. J. Tozer and N. C. Handy, J. Chem. Phys. **109**, 10180 (1998).
[16] U. von-Barth and C.-O. Almbladh, Phys. Rev. B **31**, 3231 (1985).
[17] L. A. Cole and J. P. Perdew, Phys Rev A **25**, 1265 (1982).
[18] O. A. Vydrov and G. E. Scuseria, J. Chem. Phys. **121**, 8187 (2004).
[19] J. D. Talman and W. F. Shadwick, Phys Rev A **14**, 36 (1976).
[20] F. Della Sala and A. Gorling, J. Chem. Phys. **115**, 5718 (2001).
[21] P. Mori-Sanchez, Q. Wu, and W. Yang, J. Chem. Phys. **119**, 11001 (2003).
[22] D. C. Langreth and J. P. Perdew, Sol. Stat. Comm. **17**, 1425 (1975).
[23] O. Gunnarsson and B. I. Lundqvist, Phys. Rev. B **13**, 4274 (1976).
[24] W. Yang, J. Chem. Phys. **109**, 10107 (1998).
[25] J. E. Robinson, F. Bassani, R. S. Knox, and J. R. Schreiffer, Phys. Rev. Lett. **9**, 215 (1962).
[26] R. Baer, M. Head-Gordon, and D. Neuhauser, J. Chem. Phys. **109**, 6219 (1998).
[27] R. Baer and D. Neuhauser, J. Chem. Phys. **112**, 1679 (2000).
[28] N. Troullier and J. L. Martins, Phys. Rev. B **43**, 1993 (1991).
[29] J. P. Perdew and Y. Wang, Phys. Rev. B **45**, 13244 (1992).
[30] J. M. L. Martin and P. R. Taylor, J. Chem. Phys. **102**, 8270 (1995).
[31] NIST, Reference Database No. 69 (2003).
[32] R. D. Mead, A. E. Stevens, and W. C. Lineberger, in *Gas Phase Ion Chemistry*, edited by M. T. Bowers (Academic, New York, 1984), in Vol. 3, p. 213.
[33] J. A. Pople, M. Head Gordon, D. J. Fox, K. Raghavachari, and L. A. Curtiss, J. Chem. Phys. **90**, 5622 (1989).
[34] H. Iikura, T. Tsuneda, T. Yanai, and K. Hirao, J. Chem. Phys. **115**, 3540 (2001).
[35] T. Leininger, H. Stoll, H.-J. Werner, and A. Savin, Chem. Phys. Lett. **275**, 151 (1997).
[36] J. Heyd, G. E. Scuseria, and M. Ernzerhof, J. Chem. Phys. **118**, 8207 (2003).
[37] R. Baer, T. Seideman, S. Ilani, and D. Neuhauser, J. Chem. Phys. **120**, 3387 (2003).
[38] R. Baer and N. Siam, J. Chem. Phys. **121**, 6341 (2004).
[39] R. Baer, D. Neuhauser, P. Zdanska, and N. Moiseyev, Phys Rev A **68**, 043406 (2003).